\documentstyle[aps,epsfig,twocolumn]{revtex}

\title{Quantum Dot as Spin Filter and Spin Memory}
\author{Patrik Recher, Eugene V. Sukhorukov, and Daniel Loss}
\address{ Department of Physics and Astronomy, University of Basel,\\
Klingelbergstrasse 82, CH-4056 Basel, Switzerland}

\date{\today}

\begin{document}
\twocolumn[\hsize\textwidth\columnwidth\hsize\csname
@twocolumnfalse\endcsname

\maketitle

\begin{abstract}
We consider a quantum dot in the Coulomb blockade regime weakly coupled
to current leads and show
that in the presence of a magnetic field the dot
acts as an efficient spin-filter (at the single-spin level) which produces a
spin-polarized current. Conversely, if the leads are fully spin-polarized
the up or down state of the spin
on the dot results in a large sequential or small cotunneling current,
and thus, together with ESR techniques,
the setup can be operated as a single-spin memory.

\end{abstract}


\vskip2pc]
\narrowtext

An increasing number of spin-related experiments
\cite{Prinz,Awschalom,Molenkamp,Ohno,Roukes,Ensslin}
show that the spin of the electron offers unique possibilities
for finding novel mechanisms for information processing---most notably in
quantum-confined semiconductors with unusually long spin dephasing times
approaching microseconds \cite{Awschalom}, and where spins can
be transported coherently over distances of up to 100 micrometers
\cite{Awschalom}. Besides the intrinsic interest in spin-related
phenomena, spin-based devices hold promises for future applications in
conventional \cite{Prinz} as well as in quantum computer
hardware \cite{Loss}.
One of the challenging problems for such applications is to obtain
sufficient control over the spin dynamics in  nanostructures.
In the following we address this issue and propose a quantum-dot setup
which can be either operated as a spin-filter (spin diode) to produce
spin-polarized currents or as a device to detect and
manipulate single-spin states (single-spin memory).
Both effects occur at the single-spin level and thus represent the ultimate
quantum limit of a spin-filter and spin-memory.
In both cases, we will work in the  Coulomb blockade
regime\cite{nato} and consider sequential and cotunneling processes. A
new feature of our proposal is that the spin-degeneracy
is lifted\cite{Kondo} with {\em different} Zeeman splittings in the dot and
in the leads which then
results in Coulomb
blockade peaks which are uniquely associated with a definite spin state
on the dot.

\begin{figure}
\centerline{\psfig{file=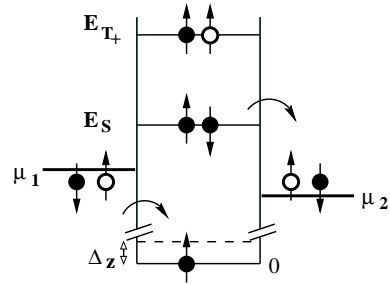,width=5cm}}
\vspace{1mm}
\caption{The energy diagram of a QD attached to two leads is
shown in the regime where the QD
contains an odd number $N$ of electrons with a top-most single electron in
the ground state ($\uparrow$ filled circle, and $E_{\uparrow}=0$).
A cotunneling process is depicted (arrows) where two possible virtual
states, singlet $E_S$ and triplet
$E_{T_+}$, are shown. The parameter $E_{S}-\mu_{1} $ can be tuned by the
gate voltage to get into the sequential tunneling regime, defined by
$\mu_{1}\ge E_{S}\ge\mu_{2}$,
where $N$ on the QD fluctuates between odd and even.
For  $N$ even, the ground state contains
a top-most singlet state with
$E_{S}<\mu_{1},\mu_{2}$.}
\label{dotfig}
\end{figure}

{\it Formalism.--\/}
Our system consists of a
quantum dot (QD) connected  to two Fermi-liquid leads
which are in equilibrium with reservoirs kept at
the chemical potentials $\mu_l$, $l=1,2$,
where outgoing currents  can be measured, see Fig.\ \ref{dotfig}.
Using a standard tunneling  Hamiltonian approach\cite{Mahan}, we write
for the full Hamiltonian $H_0+H_T$, where
$H_0=H_L+H_D$
describes the leads and the dot, with $H_D$ including the charging and
interaction energies of the dot electrons as well as the Zeeman energy
$g\mu_B B$ of their spins in the presence of an external magnetic field
${\bf B}=(0,0,B)$, where $g$ is the effective g-factor.
We concentrate first on
unpolarized lead currents and assume that the Zeeman splitting
in the leads is negligibly small compared to the one in the QD.
This can be  achieved e.g.
by using InAs for the dot ($g=15$) attached to GaAs 2DEG leads ($g=-0.44$),
or by implanting a magnetic impurity (say Mn)
inside a GaAs dot (again attached to GaAs 2DEG leads) with a strongly
enhanced electron g-factor due to exchange splitting with the
magnetic impurity \cite{OhnoScience}.
[Below we will also consider the opposite situation
with a fully spin polarized lead current, and a much smaller Zeeman
splitting on the dot.] The tunneling between leads and the QD is
described by the perturbation
$H_T=\sum_{l,k,p,s}t_{lp}c_{lks}^{\dag}d_{ps}+{\rm h.c.}$,
where  $d_{ps}$ and $c_{lks}$ annihilate electrons
with spin $s$ in the dot  and in the $l$th
lead, {\em resp.}
While the orbital $k$-dependence of the tunneling amplitude
$t_{lp}$ can be safely neglected, this is not the case
in general for the QD orbital states $p$.
{}From now on we concentrate on the Coulomb blockade (CB)
regime\cite{nato}, where  the charge in the QD,
${\hat N}=\sum_{p,s}d_{ps}^{\dag}d_{ps}$, is quantized, ${\rm Tr}\rho
{\hat N}=N$.
Next, turning to the dynamics induced by
$H_T$, we
introduce the reduced density matrix for the dot, $\rho_D={\rm Tr}_L\rho$,
where $\rho$ is the full stationary density matrix,
and ${\rm Tr}_L$ is the trace over the leads.
To describe the stationary limit, we use a standard master equation
approach \cite{nato} formulated in terms of the dot eigenstates and
eigenenergies, $H_D| n\rangle =E_n | n\rangle$, where $ n=({\bf n},N)$.
Denoting with $\rho(n)=\langle n|\rho_D| n\rangle$  the
stationary probability for the dot to be in the state $| n\rangle$, and
with  $W(n',n)$ the transition rates between $n$ and $n'$, the stationary
master equation to be solved  reads
$\sum_n\left[W(n',n)\rho(n)-W(n,n')\rho(n')\right]=0$.

The rates $W$ can be calculated in a standard ``golden rule'' approach
\cite{Sakurai} where we go up to 2nd order in $H_T$,
i.e. $W=\sum_lW_l +\sum_{l',l}W_{l'l}$,
where $W_l \propto t^2$ is the rate for a  tunneling process
of an electron from the
$l$th lead to the dot and back, while $W_{l'l}\propto t^4$ describes the
simultaneous tunneling of two electrons from the lead $l$ to the dot and
from the dot to the lead $l'$. Thus,
two regimes of  transport through the QD can be
distinguished: Sequential tunneling (ST) and cotunneling
(CT)\cite{nato,averinazarov}. The ST regime is at the degeneracy point,
where
${\hat N}$  fluctuates between $N$ and $N'=N\pm 1$, and 1st order
transitions  are allowed by
energy conservation with the explicit rates
\begin{eqnarray}
W_l(n',n)
&=&
2\pi \nu\left[
f_l(\Delta_{n'n})
 |A_{lnn'}|^2
\delta_{N',N+1}
\right. \nonumber\\
&+& \left.
[1-f_l(\Delta_{ nn'})]
|A_{ln'n}|^2
\delta_{ N',N-1}
\right],
\label{first}
\end{eqnarray}
where $\nu=\sum_k\delta(\varepsilon_F-\varepsilon_k)$ is the lead
density-of-states  per spin at the Fermi energy $\varepsilon_F$,
$f_l(\varepsilon)=[1+\exp((\varepsilon-\mu_l)/k_BT)]^{-1}$ the Fermi
function at temperature $T$,
$\Delta_{n'n}=E_{n'}-E_n$ is the level distance, and we have
introduced the matrix elements
$A_{ln'n}=\sum_{ps}t_{lps}\langle  n'|d_{ps}| n\rangle$.
In the ST regime the current through the QD  can be written as
$I_s=\pm e\sum_{n,n'}W_{2}(n',n)\rho(n)$,
where $\pm$ stands for $N'=N\mp 1$.
We emphasize that the rates $W(n,n')$ and thus the current depend
on the spin state of the dot electrons via $n,n'$.
The ST current takes a particularly
simple form if bias
$\Delta \mu=\mu_1-\mu_2>0$
and temperature are small compared to the level distance on the dot
(the case of interest here),
$\Delta \mu ,k_BT<|\Delta_{mn}|, \forall m,n$, and, thus only
the lowest energy levels  participate in the transport \cite{nato}.
The solution of the master equation gives in this case for the
ST current
\begin{equation}
I_s=\frac{e\gamma_1\gamma_2}
{\gamma_1+\gamma_2}
\left[ f_1(\Delta_{ n'n})-f_2(\Delta_{ n'n})\right],
\label{sequential}
\end{equation}
where $\gamma_l=2\pi \nu|A_{l{ nn'}}|^2$ is the  tunneling rate
through the $l$th barrier.

In the CT regime the only allowed processes are 2nd order transitions
with initial and final electron number on the QD being equal, i.e. $N=N'$,
and with the rate
\begin{eqnarray}
&& W_{l'l}({ n',n})=2\pi\nu^2\int d\varepsilon
f_l(\varepsilon)[1-f_{l'}(\varepsilon - \Delta_{ n'n})] \nonumber \\
&& \quad \times
\left|\sum_{ n_1}\frac{A_{l'{ n'n_1}}A^{*}_{l{ nn_1}}}
{\Delta_{ nn_1}+\varepsilon}+
\sum_{ n_2}\frac{A_{l'{ n_2n}}A^{*}_{l{ n_2n'}}}
{\Delta_{ n'n_2}-\varepsilon}
\right|^2\,,
\label{second}
\end{eqnarray}
where, $N_1=N+1$, and $N_2=N-1$, and thus the two terms in  Eq.\
(\ref{second}) differ by the sequence of tunneling. Our regime
of interest here is {\em elastic} CT where $E_{ n'}=E_{ n}$, which holds
for
$|\Delta_{mn}|> \Delta \mu, k_B T, \forall m\neq n$. That means the system
is always in the ground state with $\rho(n)=1$, and thus the
CT current
is given by $I_c=eW_{21}({ n,n})-eW_{12}({ n,n})$.
In particular, close to a ST resonance (but still in the CT
regime) Eq. (\ref{second}) considerably simplifies---only one
term  contributes---and for $\Delta\mu,
k_B T<|\mu \pm \Delta_{n n_i}|$,  we obtain
\begin{equation}
I_c=\frac{e}{2\pi}
\frac{\gamma_1\gamma_2\Delta\mu}
{(\mu\pm\Delta_{{ nn}_i})^2}\, ,
\label{cotunneling}
\end{equation}
where $+$ stands for $i=1$, and $-$ for $i=2$,
and $\mu=(\mu_1+\mu_2)/2$.
{}From Eqs.\ (\ref{sequential}) and (\ref{cotunneling}) it follows that
$I_s\sim \gamma_i$, while $I_c\sim \gamma_i^2$, and therefore $I_c\ll I_s$.
Thus, in the CB regime the current as a function of
$\mu$ (or gate voltage) consists of resonant ST peaks,
where ${\hat N}$ on the QD fluctuates between $N$ and
$N\pm 1$.
The peaks are separated by plateaus, where $N$ is fixed,
and where the (residual) current is due to CT.

We note that the tunneling rates
$\gamma_l$ depend on the tunneling path through
the matrix elements $A_{l{ mn}}$. In general, this can lead
to a spin-dependence of the  current,
which, however, is
difficult to measure\cite{Pustilnik}.
In contrast to this, we  will show
now that a much stronger spin dependence can come from the resonance
character of the currents
$I_s$ and $I_c$, when the position of a resonance (as function of
gate voltage) depends  on the spin orientation of the tunneling
electron.
To proceed we first
specify the energy spectrum of the QD more precisely.
In general,
the determination of the spectrum of a QD is a complicated many-electron
problem \cite{Pfannkuche}. However, it is known from
experiment\cite{Tarucha} that especially away from orbital degeneracy
points (which can be easily achieved by applying magnetic
fields\cite{Tarucha}) the spectrum is  formed  mainly by single-particle
levels, possibly slightly renormalized by exchange
interaction \cite{exchange}.

For a QD with $N$ odd  there is
one  unpaired electron in one of the two lowest energy states,
$|\uparrow\rangle$ and
$|\downarrow\rangle$,  with energies $E_{\uparrow}$ and
$E_{\downarrow}$,  which become Zeeman split due to a
magnetic field $B$,
$\Delta_z=|E_{\uparrow}-E_{\downarrow}|=\mu_B|gB|$. Let us assume that
$|\uparrow\rangle$ is the ground state, and set $E_{\uparrow}=0$
for convenience.
For $N$ even, the two topmost electrons (with the same orbital wave
function), form a spin singlet,
$(|\uparrow\downarrow\rangle-|\downarrow\uparrow\rangle)/\sqrt{2}$,
with energy $E_S$. This is the ground state, while the other states,
such as three triplets $|T_{+}\rangle=|\uparrow\uparrow\rangle$,
$|T_{-}\rangle=|\downarrow\downarrow\rangle$, and
$|T_{0}\rangle=(|\uparrow\downarrow\rangle
+|\downarrow\uparrow\rangle)/\sqrt{2}$
with  energies $E_{T_{\pm}}$ and $E_{T_0}$ are excited states,
because of higher (mostly) single-particle orbital energy.

\begin{figure}
\centerline{\psfig{file=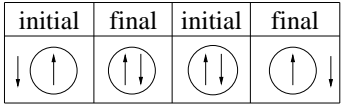,width=4.5cm}}
\vspace{1mm}
\caption{The only allowed processes  for charge transport
through the dot in
the
ST regime at the odd-to-even transition. A spin-down
electron tunnels from lead 1 to the dot forming a singlet
and tunnels out
again into lead 2. Tunneling of spin-up electrons  into
the dot is forbidden
by energy conservation since this
process involves excited states. The resulting current, $I_s(\downarrow)$,
is spin-polarized.}
\label{spintabl1}
\end{figure}

{\it  Sequential tunneling.\/}
First we consider the ST peak, which separates
two plateaus with $N$ and $N+1$ electrons on the dot, where $N$
is odd (odd-to-even transition). In the regime
$E_{T_+}-E_S,\Delta_{z}>\Delta\mu$, $k_{B}T$, only ground-state
transitions are allowed by energy conservation.
The tunneling of spin-up electrons is  blocked by energy conservation,
\ i.e. $I_s(\uparrow)=0$,
because
it involves excited states $|T_{+}\rangle$ and $|\downarrow\rangle$.
The only possible process is tunneling of spin-down electrons as
shown in Fig.\ \ref{spintabl1}, which leads to a {\it spin-polarized}
current, $I_s(\downarrow)$,
given by Eq.\ (\ref{sequential}),
where $\Delta_{n'n}=E_S>0$ (since $E_{\uparrow}=0$).
Thus, we have
\begin{eqnarray}
&& I_s(\downarrow)/I_0=\theta(\mu_1-E_S)-\theta(\mu_2-E_S), \quad
k_B T<\Delta\mu ,
\label{small-T} \\
&& I_s(\downarrow)/I_0=
\frac{\Delta\mu}{4k_BT}\cosh^{-2}\left[\frac{E_S-\mu}{2k_BT}\right],
\quad k_BT >\Delta\mu,
\label{large-T}
\end{eqnarray}
where $I_0=e\gamma_1\gamma_2/(\gamma_1+\gamma_2)$. Hence, in the specified
regime
the dot acts as spin filter through which only
spin-down electrons can pass \cite{Kramernote}.

Next, we consider the ST peak at the
transition from even to odd, i.e.\
when $N$ is even. Then the current is given by  Eq.\ (\ref{sequential})
with $\Delta_{ n'n}=-E_S>0$. The spin-down current is now blocked,
$I_s({\downarrow})=0$, while spin-up electrons can pass through the QD, with
the
current
$I_s({\uparrow})$ given  by (\ref{small-T}) and (\ref{large-T}), where
$E_S$ has to
be replaced by
$-E_S$. Because this case is very similar to the previous one
with $\downarrow$ replaced by $\uparrow$, we shall
concentrate on the odd-to-even  transition only.
Next, we will
demonstrate that although CT processes
can in general lead to a leakage of unwanted current,
this turns out to be a minor effect, and spin filtering
works also in the CT regime.

\begin{figure}
\centerline{\psfig{file=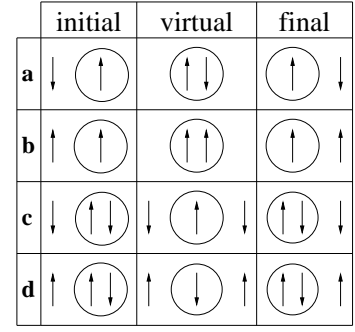,width=4.5cm}}
\vspace{1mm}
\caption{(a)  and (b)  are the main processes in the
cotunneling regime with $N$ odd  when inelastic processes and
processes where the dot is not in the ground state are suppressed by the
Zeeman energy $\Delta_{z}$. Only the leading virtual transitions are
shown. (c)  and (d)
visualize the leading cotunneling processes for $N$ even. Here, other
processes are suppressed by the energy difference
between singlet  and triplet, $E_{T_{+}}-E_{S}$ .}
\label{spintabl}
\end{figure}

{\it Cotunneling.\/}
Above or below a ST resonance, i.e. when
$E_{S}>\mu_{1,2}$ or
$E_{S}<\mu_{1,2}$,
the system is in the CT regime.
Close to this peak the main contribution to the transport is due to two
processes (a) and (c) {see Fig.\ \ref{spintabl}}, where the energy deficit
of
the virtual states,
$|\mu -E_S|$, is minimal. According to Eq.\ (\ref{cotunneling})
we have
\begin{equation}
I_c(\downarrow)=\frac{e}{2\pi}\frac{\gamma_1\gamma_2\Delta\mu}{(\mu-E_S)^2}.
\label{main}
\end{equation}
Thus, we expect the spin filtering of down electrons to work even in the
CT regime close to the resonance.
However, there are additional CT processes, (b) and (d), which
involve tunneling of spin-up electrons and lead to a leakage of
up-spin. If $N$ is odd (below the resonance),
the dot is initially in its ground state
$(\uparrow)$,  and an incoming spin-up electron
can only form  a  virtual triplet state $|T_{+}\rangle$ (process (b) in
Fig. \ref{spintabl}). This process contributes to the rate
(\ref{sequential})
with an energy deficit $E_{T_+}-\mu$, so that for the efficiency
of spin filtering [defined as $I(\downarrow)/I(\uparrow)$]
we obtain in this regime,
\begin{equation}
I_c(\downarrow)/I_c(\uparrow)\sim
\left(1+\frac{E_{T_+}-E_S}{E_S-\mu}\right)^2, \quad \mbox{$N$  odd.}
\label{efficiency1}
\end{equation}

Above the resonance, i.e.\ when $N$ is even and the ground state
is the spin-singlet $|S\rangle$, the tunneling of spin-up
electrons occurs via the virtual spin-down state (process (d) in
Fig. \ref{spintabl}) with an energy deficit $\Delta_z+\mu-E_S$,
which has to be compared to the energy deficit $\mu-E_S$ of the main
process (c). Thus, we obtain for the efficiency of the spin filtering
in the CT regime
\begin{equation}
I_c(\downarrow)/I_c(\uparrow)\sim\left(1+\frac{\Delta_z}
{\mu-E_S}\right)^2,\
\quad
\mbox{$N$  even.}
\label{efficiency2}
\end{equation}

We see that in both cases, above and below the resonance, the efficiency
can be made large by tuning the gate voltage to the resonance, i.e. \
$|\mu -E_S|\to 0$. Eventually, the system goes to the ST regime,
$|\mu - E_S|\lesssim k_BT,\Delta\mu$. Now, using Eqs.\
(\ref{cotunneling}), (\ref{small-T}), and (\ref{large-T}) we can estimate
the efficiency of spin filtering in the ST regime,
\begin{equation}
I_s(\downarrow)/I_c(\uparrow)\sim
\frac{\Delta_z^2}{(\gamma_1+\gamma_2)\max\{k_BT,\Delta\mu\}},
\label{efficiency3}
\end{equation}
 where we have assumed
 that $\Delta_z<|E_{T_+}-E_S|$.
In the ST regime considered here
we have $\gamma_i< k_{B}T,\Delta\mu$ \cite{nato}.
Therefore,  if the requirement $k_{B}T,\Delta\mu<\Delta_z$ is satisfied,
filtering of spin-down electrons in the  ST regime is
very effecient.
In the quantum regime,
$\gamma_i> k_{B}T,\Delta\mu$, tunneling occurs as Breit-Wigner
resonance\cite{nato}, and $\max\{k_BT,\Delta\mu\}$ in Eq.\
(\ref{efficiency3}) has to be replaced by $\gamma_i$.
Finally, we note that the spin polarization of the
transmitted current oscillates between up and down as we change the number
of dot electrons $N$ one by one.

The functionality of the spin filter can  be tested e.g.
with the use of a p-i-n diode \cite{Molenkamp,Ohno} which is placed in
the outgoing lead 2. Via  excitonic
photoluminescence, the diode transforms the spin polarized
electrons (entering lead 2) into correspondingly circularly polarized
photons which can then be detected.

{\it Spin memory.\/}
We consider now the opposite case where the current in the
leads is fully spin polarized.
Recent experiments have demonstrated  that fully
spin-polarized carriers can be tunnel-injected from a spin-polarized
magnetic semiconductor (III-V or II-VI materials) with large effective
g-factor into an unpolarized GaAs system \cite{Ohno,Molenkamp}.
Another possibility would be to work in the quantum Hall regime
where spin-polarized edge states are coupled to a quantum
dot\cite{Sachrajda}.
To be specific, we consider the case where $E_{T_{+}}-E_s +\Delta_z > \Delta
\mu,
k_BT$ with $E_{T_{+}}>E_s$  ($\Delta_z >  k_BT$ is not necessary as long
as the spin relaxation time is longer than the measurement time, see
below). We assume that the spin polarization of both leads is, say,
up and $N$ is odd.
There are now two cases for the current, $I^{\uparrow}$ or $I^{\downarrow}$,
corresponding to a spin up or down on the QD.
 First, we assume the QD to be in the ground state
with its topmost electron-spin pointing up. According to previous
analysis [see paragraph before Eqs. (\ref{small-T},\ref{large-T})], the
ST current vanishes, i.e.  $I_s^{\uparrow}=0$,
since the tunneling into the level $E_{T_{+}}$ (and higher levels) is
blocked by energy conservation, while the tunneling into $E_s$ is
blocked by spin conservation (the leads can provide and take up only
electrons
with spin up).
However, there is again a small CT current, $I_c^{\uparrow}$,
which is given by Eq. (\ref{main}). Now we compare this to the second case
where
the topmost dot-spin is down with additional Zeeman energy $\Delta_z>0$.
Here, the ST current $I_s^{\downarrow}$ is finite, and
given by Eqs. (\ref{small-T}, \ref{large-T}) with $E_s$ replaced by $E_s-
\Delta_z$. In the limit $E_s> \Delta_z$, we
get $I_s^{\downarrow}=I_s({\downarrow})$, and thus the ratio
$I_s^{\downarrow}/I_c^{\uparrow}$ is again given by Eq.
(\ref{efficiency3}). Hence, we see that the dot with its spin up transmits
a much smaller current than the dot with spin down. This fact allows
the read-out of the spin-state of the (topmost) dot-electron by
comparing the measured currents. Furthermore, the spin state of the QD can
be
changed (``read-in") by ESR techniques, i.e. by applying a pulse
of an ac magnetic field (perpendicular to {\bf B}) with resonance
frequency $\omega=\Delta_z$\cite{Burkard}. Thus the proposed
setup functions
as a single-spin memory with read-in and read-out capabilities,
the relaxation
time of the memory given by the spin relaxation time $\tau_S$
on the QD (which can be expected to exceed 100's of
nanoseconds\cite{Awschalom}). We note that this $\tau_S$ can be easily
measured since it is the time during which
$I_s^{\downarrow}$ is finite before it strongly reduces to
$I_c^{\uparrow}$.
Finally, this scheme can be
upscaled: In an array of such QDs where each dot separately is attached
to in- and outgoing leads (for read-out) we can switch the spin-state of
each dot individually by locally controlling the Zeeman splitting
$\Delta_z$. This can be done
\cite{Loss} e.g. by applying a gate voltage on a particular dot that
pushes the wave function of the dot-electrons into a region of, say,
higher effective g-factor (the induced level shift in the QD can be
compensated for by the chemical potentials).

In conclusion, we have shown that  quantum dots in the Coulomb blockade
regime and attached to leads can be operated as efficient spin filters
at the single-spin level.
Conversely, if the leads are spin-polarized the spin state of the
quantum dot can be read-out by a traversing current which is (nearly)
blocked for one spin state while it is unblocked for the opposite spin
state.

{\it Acknowledgements.}
This work has been supported
by the Swiss NSF.


\begin{references}
\bibitem{Kondo} Note that by breaking the spin-degeneracy
Kondo effects are excluded.
\bibitem{exchange}
When  exchange interactions become dominant (e.g. for $N$ very small), our
analysis of course still applies, since all we require is a sufficiently
large energy splitting $\Delta_{n'n}$ between singlet and triplet.
\bibitem{Kramernote}
We mention that ``spin-blockade" effects based on spin-selection
rules have been considered before\protect\cite{Kramer}. However,
these effects are different from the ones found here, in particular,
they do not lead to spin-polarized currents, they
appear  only in the non-linear
voltage regime for $\Delta S=1/2$, and they vanish with increasing
magnetic field\cite{Kramer}---all in contrast to the case considered
here.
\end{references}
\end{document}